\documentclass{article}
\usepackage[a4paper]{geometry}
\usepackage{amsmath}

\input{tcilatex}
\begin{document}

\title{Gaussian wave packet solution of the Schrodinger equation in the
presence of a time-dependent linear potential\ }
\author{M. Maamache and Y. Saadi \\
\\
\textit{Laboratoire de Physique Quantique et Syst\`{e}mes Dynamiques,}\\
\textit{{Facult\'{e} des Sciences,Universit\'{e} Ferhat Abbas de S\'{e}tif},
S\'{e}tif 19000, Algeria}}
\date{}
\maketitle

\begin{abstract}
We argue that the way to get the general solution of a Schrodinger equation
in the presence of a time-dependent linear potential based on the
Lewis-Riesenfeld framework is to use a Hermitian linear invariant operator.
We demonstrate that the linear invariant proposed in $p$ and $q$ is an
Hermitian operator which has the Gaussian wave packet as its eigenfunction.

\end{abstract}

The time evolution of a quantum system subject to a spatially uniform,
time-dependent force has attracted considerable interest lately. The exact
propagator for this system has long been known [1], as have a set of exact
solutions (the Volkov solutions) [2-3].\ Recent work [4--13] focuses further
on exact solutions and their properties. First, Guedes [5] obtained, by
means of the invariant operator method introduced by Lewis and Riesenfeld
(L-R) [14], a special solution for the time-dependent linear potential. The
idea is that any operator satisfying the quantum Liouville-von Neumann
equation provides its eigenstate as a solution of the time-dependent
Schrodinger equation up to a time dependent phase factor. Later on, Feng [6]
followed a method based on spatiotemporal transformations of the Schrodinger
equation to get the plane-wave-type and the Airy-packet solutions where the
one in Ref. [5] constitutes a particular case which corresponds to the
so-called \textquotedblleft standing\textquotedblright\ particle case in a
linear potential [6].

However, Bekkar et al. [8] pointed out that the Airy-packet solution is in
fact only a superposition of the plane-wave-type solution. In his Comment
[7], Bauer explained that the solution found by Guedes [5] is simply a
special case of the Volkov solution, with a zero wave vector $k$, to the
time dependent Schrodinger equation describing a nonrelativistic charged
particle moving in an electromagnetic field. Dunkel and Trigger [10]
considered the initial minimum-uncertainty Gaussian \ for a sinusoidally
time-dependent linear potential. Bowman [11] investigated the time evolution
of the general quantum state for the time-dependent linear system, which was
shown to be that of a free-particle state, plus an overall motion arising
from the classical force . Sang Pyo Kim [12] has shown that a charged free
particle in a constant and/or oscillating electric field has a bounded
Gaussian wave packet, which is a coherent state of a one-parameter and
time-dependent ground state for the free-particle Hamiltonian. From the
test-function method, Gengbiao Lu et al.[13] constructed an exact n
wave-packet-train (n GWPT) solution , whose center moves acceleratively
along the corresponding classical trajectory.

Very recently, Luan et al. [9] have reexamined the linear invariant proposed
by Guedes [5] and indicated that besides the solutions described in Refs.
[5-8], a Gaussian wave packet eigenfunction solution can naturally be
derived from the LR method if a non-Hermitian linear invariant is used, and
they argue that this solution was ruled out before, because the authors in
Refs. [ 5-8] assumed in advance the linear L-R invariant $%
I(t)=A(t)p+B(t)x+C(t)$ be a Hermitian operator.

The authors [9] stated that in Ref.[5, 8] , the Gaussian wave packet
eigenfunction solution was ruled out , because the linear LR invariant is
supposed to be Hermitian operator, and the $B$ parameter was set to zero for
hermiticity reason. In fact this is not the reason for setting \ $B=0$, in
Ref. [8], it was meant to justify the $B=0$ imposed in Ref. [5]. We will
show and we will correct in the present paper, that if $B$ is different from
zero ($B\neq 0$), we still obtain Gaussian wave packet (GWP) solutions. In
fact, we could find all the results using \ the Hermitian invariant operator
linear in $p$ and $q$ with any conditions imposed to get physically
acceptable solutions as was done in Luan et al. paper [9]. In the latter
approach, the authors obtained solutions labelled by complex parameters and
complex eigenvalues, this is ambiguous.

We recall that according to the theory of Lewis and Riesenfeld [14], an
invariant is an operator that must necessarily satisfy three requirements:
1. It is Hermitian. 2. It satisfies the Von Neumann equation. 3. Its
eigenvalues are real and time-independent (we signal that the Hermiticity of
the invariant is one of the essential conditions which makes the eigenvalues
of the invariant time-independent). Furthermore, any invariant satisfying
these three requirements leads to a complete set of solutions of the
corresponding Schrodinger equation. So, a conventional solution is
constructed as a linear combination of these solutions.

\bigskip The problem is to find the solutions of the Schrodinger equation,

\begin{equation}
i\hbar \frac{\partial }{\partial t}\psi (x,t)=H\text{ }\psi (x,t)
\end{equation}

for the Hamiltonian

\begin{equation}
H\left( x,p,t\right) =\frac{1}{2m}p^{2}-F(t)x,
\end{equation}%
where $F(t)$ is a time-dependent function.

According to the theory of Lewis and Riesenfeld [14], a solution of the Schro%
\"{}%
dinger equation with a time-dependent Hamiltonian is easily found if a
nontrivial Hermitian operator $I(t)$ exists and satisfies the invariant
equation

\begin{equation}
\frac{dI}{dt}=\frac{\partial I}{\partial t}+\frac{1}{i\hbar }\left[ I,H%
\right] =0.
\end{equation}

\bigskip Indeed, this equation is equivalent to saying that if $\varphi
_{\lambda }(x,t)$ is an eigenfunction of $I(t)$ with a time-independent
eigenvalue $\lambda $, we can find a solution of the Schrodinger equation in
the form $\psi _{\lambda }(x,t)=\exp \left[ i\alpha _{\lambda }\left(
t\right) \right] \varphi _{\lambda }(x,t)$ where $\alpha _{\lambda }\left(
t\right) $ satisfies the eigenvalue equation for the Schrodinger operator,

\begin{equation}
\hbar \dot{\alpha}_{\lambda }\left( t\right) \varphi _{\lambda }=i\left[
\hbar \frac{\partial }{\partial t}-H\right] \varphi _{\lambda }.
\end{equation}%
It turns out that the time-dependent invariant operator takes the linear
form [9]

\begin{equation}
I(t)=\text{ }A(t)p+B(t)x+C(t).
\end{equation}

\bigskip The invariant equation is satisfied if the time-dependent
coefficients are such that

\begin{equation}
A\left( t\right) =A_{0}-\frac{B_{0}}{m}t,\text{ \ \ \ \ \ \ \ \ \ \ \ \ \ }%
B(t)=B_{0},
\end{equation}

\begin{equation}
C\left( t\right) =C_{0}-A_{0}\int_{0}^{t}F(\tau )d\tau +\frac{B_{0}}{m}%
\int_{0}^{t}F(\tau )\tau d\tau ,
\end{equation}
where $A_{0}$, $B_{0}$, $C_{0}$ are arbitrary real constants.

\bigskip The eigenstates of $I(t)$ corresponding to time-independent
eigenvalues are the solutions of the equation

\begin{equation}
I(t)\varphi _{\lambda }(x,t)=\lambda \text{ }\varphi _{\lambda }(x,t).
\end{equation}

It is easy to see that the solutions of Eq. (8) are of the form

\begin{equation}
\varphi _{\lambda }(x,t)=\exp \left\{ \frac{i}{\hbar }\left[ \frac{2\left(
\lambda -C\left( t\right) \right) x-B_{0}x^{2}}{2A\left( t\right) }\right]
\right\} .
\end{equation}

\bigskip Substituting Eq. (9) into Eq. (4) and accomplishing the
integration, we obtain

\begin{eqnarray}
\alpha _{\lambda }\left( t\right) &=&\alpha _{\lambda }\left( 0\right)
-\int_{0}^{t}\left\{ \frac{\left( \lambda -C\left( \tau \right) \right) ^{2}%
}{2m\hbar A\left( \tau \right) ^{2}}+\frac{iB_{0}}{2mA\left( \tau \right) }%
\right\} d\tau  \notag \\
&=&\alpha _{\lambda }\left( 0\right) -\int_{0}^{t}\left\{ \frac{\left(
\lambda -C\left( \tau \right) \right) ^{2}}{2m\hbar A\left( \tau \right) ^{2}%
}\right\} d\tau -i\text{Ln}\left( \sqrt{\frac{A_{0}}{A\left( t\right) }}%
\right) ,
\end{eqnarray}%
note that the logarithmic term goes downstairs as a time-dependent
normalization factor in $\psi _{\lambda }(x,t).$

Therefore the \ physical orthogonal wave functions $\psi _{\lambda }(x,t)$
solutions of the Schrodinger equation (1 ) are given by

\begin{equation}
\psi _{\lambda }(x,t)=\sqrt{\frac{A_{0}}{A\left( t\right) }}\exp \left\{
-i\int_{0}^{t}\frac{\left( \lambda -C\left( t^{\prime }\right) \right) ^{2}}{%
2\hbar mA\left( t^{\prime }\right) ^{2}}dt^{\prime }+\right\} \exp \left\{ 
\frac{i}{\hbar }\left[ \frac{2\left( \lambda -C\left( t\right) \right)
x-B_{0}x^{2}}{2A\left( t\right) }\right] \right\} .
\end{equation}%
It is easy to verify that

\begin{equation}
\left\langle \psi _{\lambda }\mid \psi _{\lambda ^{\prime }}\right\rangle
=\delta (\lambda -\lambda ^{\prime }).
\end{equation}

Furethermore, the evolution of the general Schrodinger state can be writen as

\begin{equation}
\Psi \left( x,t\right) =\int_{-\infty }^{+\infty }g\left( \lambda \right)
\psi _{\lambda }\left( x,t\right) d\lambda ,
\end{equation}%
where $g\left( \lambda \right) $ is a weight function which determines the
state of the system such that $\Psi \left( x,t\right) $ is square integrable
i.e. $\int \mid \Psi \left( x,t\right) $ $\mid dx$ is time-independent
finite constant.

In Ref. [8] , it has been shown that the choice

\begin{equation}
g\left( \lambda \right) =\exp \left( \frac{i\lambda ^{3}}{3}\right)
\end{equation}%
leads to Airy function solutions.

Here we shall show that one obtains a general Gaussian wave-packet solution
by choosing weight function as a gaussian too, 
\begin{equation}
g\left( \lambda \right) =\sqrt{\frac{\sqrt{a}}{\hbar A_{0}\pi \sqrt{2\pi }}}%
\exp \left( -a\lambda ^{2}\right) ,
\end{equation}
$\ $where $a$ is a positive real constant.

Substituting Eq. (11) and \ (15) into Eq. (13) and accomplishing the
integration, we obtain\ the normalized Gaussian solution as

\begin{eqnarray}
\Psi \left( x,t\right) &=&\sqrt{\frac{\sqrt{a}}{\hbar A\left( t\right) \sqrt{%
2\pi }\left( a+i\int_{0}^{t}\frac{1}{2\hbar mA\left( t^{\prime }\right) ^{2}}%
dt^{\prime }\right) }}\exp \left\{ -i\int_{0}^{t}\frac{C\left( t^{\prime
}\right) ^{2}}{2\hbar mA\left( t^{\prime }\right) ^{2}}dt^{\prime }\right\} 
\notag \\
&&\exp \left\{ \frac{i}{\hbar }\left[ \frac{-2C\left( t\right) x-B_{0}x^{2}}{%
2A\left( t\right) }\right] \right\} \exp \left\{ \frac{-\left[ \int_{0}^{t}%
\frac{C\left( t^{\prime }\right) }{\hbar mA\left( t^{\prime }\right) ^{2}}%
dt^{\prime }+\frac{x}{\hbar A\left( t\right) }\right] ^{2}}{\left(
4a+i\int_{0}^{t}\frac{2}{\hbar mA\left( t^{\prime }\right) ^{2}}dt^{\prime
}\right) }\right\}
\end{eqnarray}

\bigskip Let us now evaluate the mean value of $x$, $p$ and the quantum
coordinate and momentum fluctuation in the state $\Psi \left( x,t\right) .$
After some algebra we find that \ 

\begin{equation}
\left\langle x\right\rangle =\left\langle \Psi \left( t\right) \left\vert
x\right\vert \Psi \left( t\right) \right\rangle =-A\left( t\right)
\int_{0}^{t}\frac{C\left( t^{\prime }\right) }{mA\left( t^{\prime }\right)
^{2}}dt^{\prime },
\end{equation}%
which is nothing but the classical position $x_{c}(t)$, and

\begin{equation}
\left\langle p\right\rangle =\left\langle \Psi \left( t\right) \left\vert
p\right\vert \Psi \left( t\right) \right\rangle =\frac{-C\left( t\right) }{%
A\left( t\right) }+\int_{0}^{t}\frac{B_{0}C\left( t^{\prime }\right) }{%
mA\left( t^{\prime }\right) ^{2}}dt^{\prime }
\end{equation}
is a classical momentum $p_{c}(t).$ The position uncertainty

\begin{equation}
\Delta x=\sqrt{\left\langle x^{2}\right\rangle -\left\langle x\right\rangle
^{2}}=\hbar A\left( t\right) \sqrt{\frac{\left( a^{2}+\left( \int_{0}^{t}%
\frac{1}{2\hbar mA\left( t^{\prime }\right) ^{2}}dt^{\prime }\right)
^{2}\right) }{a}},
\end{equation}%
\ and the momentum uncertainty

\begin{eqnarray}
\Delta p &=&\sqrt{\left\langle p^{2}\right\rangle -\left\langle
p\right\rangle ^{2}}  \notag \\
&=&\frac{1}{\Delta x}\sqrt{\frac{\hbar ^{2}}{4\ }+\left[ \int_{0}^{t}\frac{1%
}{4a\ mA\left( t^{\prime }\right) ^{2}}dt^{\prime }-\frac{\hbar
^{2}B_{0}A\left( t\right) }{a}\left( a^{2}+\left( \int_{0}^{t}\frac{1}{%
2\hbar mA\left( t^{\prime }\right) ^{2}}dt^{\prime }\right) ^{2}\right) %
\right] ^{2},}
\end{eqnarray}
leads to the uncertainty relation 
\begin{equation}
\Delta p\Delta x=\sqrt{\frac{\hbar ^{2}}{4\ }+\left[ \int_{0}^{t}\frac{1}{%
4a\ mA\left( t^{\prime }\right) ^{2}}dt^{\prime }-\frac{\hbar
^{2}B_{0}A\left( t\right) }{a}\left( a^{2}+\left( \int_{0}^{t}\frac{1}{%
2\hbar mA\left( t^{\prime }\right) ^{2}}dt^{\prime }\right) ^{2}\right) %
\right] ^{2}}\geq \frac{\hbar }{2}.
\end{equation}

\bigskip

We can rewrite (16) as follows \ 

\begin{eqnarray}
\Psi \left( x,t\right) &=&\sqrt{\frac{\hbar A\left( t\right) \left(
a-i\int_{0}^{t}\frac{1}{2\hbar mA\left( t^{\prime }\right) ^{2}}dt^{\prime
}\right) }{\sqrt{2\pi a}\Delta x^{2}}}\exp \left\{ -i\int_{0}^{t}\frac{%
C\left( t^{\prime }\right) ^{2}}{2\hbar mA\left( t^{\prime }\right) ^{2}}%
dt^{\prime }\right\}  \notag \\
&&\exp \left\{ \frac{i}{\hbar }\left[ \frac{-2C\left( t\right) x-B_{0}x^{2}}{%
2A\left( t\right) }\right] \right\} \exp \left\{ \frac{-\left(
a-i\int_{0}^{t}\frac{1}{2\hbar mA\left( t^{\prime }\right) ^{2}}dt^{\prime
}\right) \left[ x-\left\langle x\right\rangle \right] ^{2}}{4a\Delta x^{2}}%
\right\} .
\end{eqnarray}

Morever, the time-dependent probability density associated with this
Gaussian wave packet is Gaussian for all times

\begin{equation}
\left\vert \Psi \left( x,t\right) \right\vert ^{2}=\frac{1}{\sqrt{2\pi }%
\Delta x}\exp -\left\{ \frac{\left( x-\left\langle x\right\rangle \right)
^{2}}{2\Delta x^{2}}\right\} ,
\end{equation}%
we see that $\Delta x$ represents the width of the wave packet at time $t$.
It is also readily verified tat the time-dependent probability density is
conserved

\begin{equation}
\int_{-\infty }^{\infty }\left\vert \Psi \left( x,t\right) \right\vert
^{2}dx=1
\end{equation}

Equations (22) and (23) describe a Gaussian wave packet that is centered at $%
x=$ $\left\langle x\right\rangle $ whose width $\Delta x(t)$ varies with
time. So, during time $t$, the packet's center has moved from $x=0$ to $%
x=-A\left( t\right) \int_{0}^{t}\frac{C\left( t^{\prime }\right) }{mA\left(
t^{\prime }\right) ^{2}}dt^{\prime }$ and its width has expanded from $%
\Delta x_{0}=\hbar A_{0}\sqrt{a}$ to $\Delta x(t)=\Delta x_{0}\frac{A\left(
t\right) }{A_{0}}\sqrt{1+\left( \int_{0}^{t}\frac{1}{2\hbar maA\left(
t^{\prime }\right) ^{2}}dt^{\prime }\right) ^{2}}$. The wave packet
therefore undergoes a distortion; although it remains Gaussian, its width
broadens with time whereas its height, $\frac{1}{\sqrt{2\pi }\Delta x}$,
decreases with time. \ Further, it should be noted that the width of the
Gaussian packet does not depend on the external force $F(t)$. Thus the shape
of the wave packet is not changed by the external force. This means that the
external force $F(t)$ acts uniformly in the wave packet.

Acknowledgments:

We wish to thank Professor A. Layadi for his help.

References

[1] R. Feynman and A. Hibbs, Quantum Mechanics and Path Integrals (New York:
McGraw-Hill), (1965).

[2] W. Gordon, Z. Phys. 40, 117 (1926).

[3] D. Volkov, Z. Phys. 94, 250 \ (1935).

[4] A. Rau and K. Unnikrishnan, Phys. Lett. A 222, 304 (1999).

[5] I. Guedes, Phys. Rev. A 63, 034102 (2001).

[6] M. Feng, Phys. Rev. A 64, 034101 (2001).

[7] J. Bauer, Phys. Rev. A 65, 036101 (2002).

[8] H. Bekkar, F. Benamira, and M. Maamache, Phys. Rev. A 68, 016101 (2003).

[9] Pi-Gang Luan and Chi-Shung Tang, Phys. Rev. A 71, 014101 (2005).

[10] J. Dunkel and S. A. Trigger, Phys. Rev.A 71, 052102 (2005).

[11] G.E. Bowman, J. Phys. A. 39, 157 (2006).

[12] S. P. Kim, J. Korean Phys. Soc. 44, 464 (2006).

[13] Gengbiao Lu, Wenhua Hai and Lihua Cai, Phys. Lett. A 357, 181 (2006).

[14] H. R. Lewis, Jr. and W. B. Reisenfeld, J. Math. Phys. 10, 1458 (1969).

\bigskip

\end{document}